\input epsf
\documentstyle[twoside,fleqn,npb,epsfig,eqnfix]{article}
   \newcommand{\be}{\begin{equation}}
   \newcommand{\ee}{\end{equation}}
   \def\e{\epsilon}
   \def\half{{\textstyle{1\over 2}}}
   \def\P{I\!\!P}
   \newcommand{\beqa} {\begin{eqnarray}}
   \font \fiverm=cmr7
   \newcommand{\eeqa} {\end{eqnarray}}
   \newcommand{\nn} {\nonumber}
   \def\bu{{$\bullet ~$}}

\newcommand{\AmS}{{\protect\the\textfont2
  A\kern-.1667em\lower.5ex\hbox{M}\kern-.125emS}}

\title{Pomeron physics: an update}

\author{P V Landshoff\address{DAMTP, Cambridge University \\
        Cambridge CB3 9EW, England \\
        pvl@damtp.cam.ac.uk}%
        \thanks{This research is supported in part by the EU Programme
``Training and Mobility of Researchers", Network
``Quantum Chromodynamics and the Deep Structure of
Elementary Particles'' (contract FMRX-CT98-0194),
and by PPARC}}

\begin{document}

\begin{abstract}
Key issues in pomeron physics include whether the hard and soft
pomerons are distinct objects, and whether the hard pomeron is already
present in amplitudes at $Q^2=0$. It is urgent to learn how to combine
perturbative and nonperturbative concepts, and to construct a sound
theory of perturbative evolution at small $x$. Other questions are
whether screening corrections are small, and gap survival probabilities
large. Finally, do diffractive processes present a good way to
discover the Higgs? 
\end{abstract}

\maketitle

\section{INTRODUCTION}

The idea of the 
pomeron is some 40 years old. Already in its early days it was hugely
successful in explaining and correlating a wide variety of reactions. 
In those days, the theoretical emphasis was on matrix elements being
analytic functions of all their variables: $s,t,\ell,\dots$ (where
$\ell$ represents orbital angular momentum). It was recognised that
the positions of the singularities of the amplitudes depend only
on the masses of the unphysical particles.

Nowadays the emphasis is more on perturbative QCD. Confinement 
is largely ignored, with the result that the singularities of
the amplitudes do not depend on physical particle
masses. It is urgent to learn how to reconcile the two approaches.
They are not to be regarded a rivals: rather, we have to learn how
to include nonperturbative effects into the perturbative formalism
sufficiently to remove any conflicts with the fundamental properties 
of scattering amplitudes known to follow from from basic principles 
such as unitarity.

\section{Total cross sections}

\begin{figure}[t]
\begin{center}
\vskip -1truemm
\epsfxsize=.45\textwidth\epsfbox[98 500 345 770]{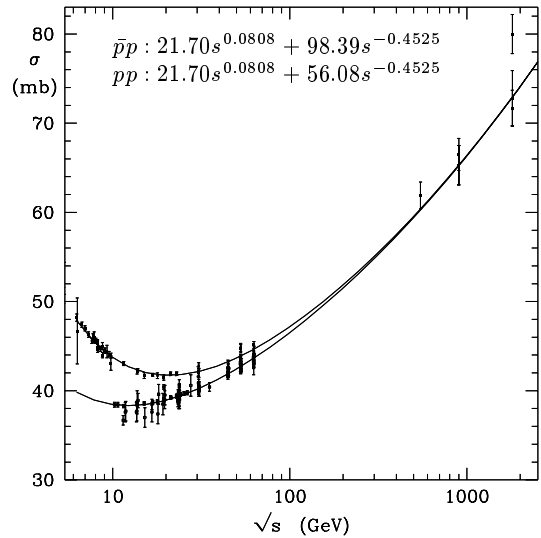}
\end{center}
\vskip -16truemm
\caption{$pp$ and $\bar pp$ total cross-sections}
\label{SIGTOT1}
\vskip -7truemm
\end{figure}
\begin{figure}[t]
\begin{center}
\vspace{-7mm}
\epsfxsize=.45\textwidth\epsfbox[98 500 340 770]{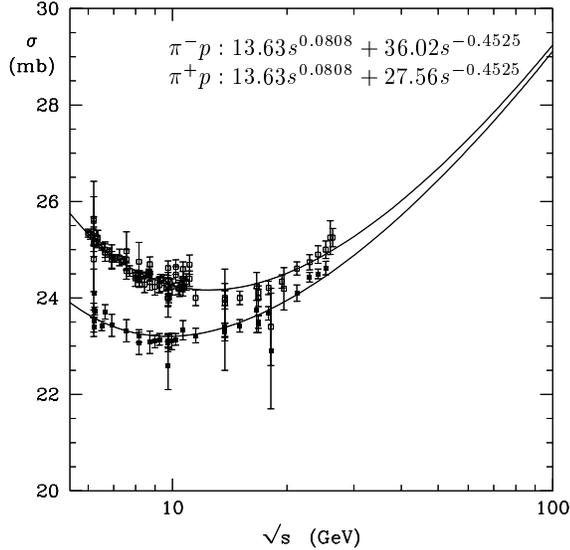}
\end{center}
\vskip -16truemm
\caption{$\pi^+p$ and $\pi^-p$ total cross-sections}
\label{SIGTOT2}
\vskip -7truemm
\end{figure}

Figure \ref{SIGTOT1} shows data for the $pp$ and $\bar pp$ total cross-sections,
together with fits\cite{DL92} that include the exchange of the soft pomeron and
the $\omega,\rho,f_2$ and $a_2$ families of particles. As is evident, there
is a conflict between the two Tevatron measurements\cite{Amo89,CDF94}. 
I do not think it is
sensible to produce a fit that goes half way between the two points. If the
upper (CDF) measurement turns out to be correct, rather than the lower
(E710) point, it could be an indication of a new and interesting
contribution setting in at high energy, which could be confirmed at the
LHC. There are some high-energy cosmic-ray data, but as Matthiae has
explained at this meeting, extracting the $pp$ cross section from air-%
shower data is subject to very considerable uncertainties\cite{EGLT98}.

\begin{figure}[t]
\begin{center}
\vskip -5truemm
\epsfxsize=0.42\textwidth\epsfbox[55 540 360 765]{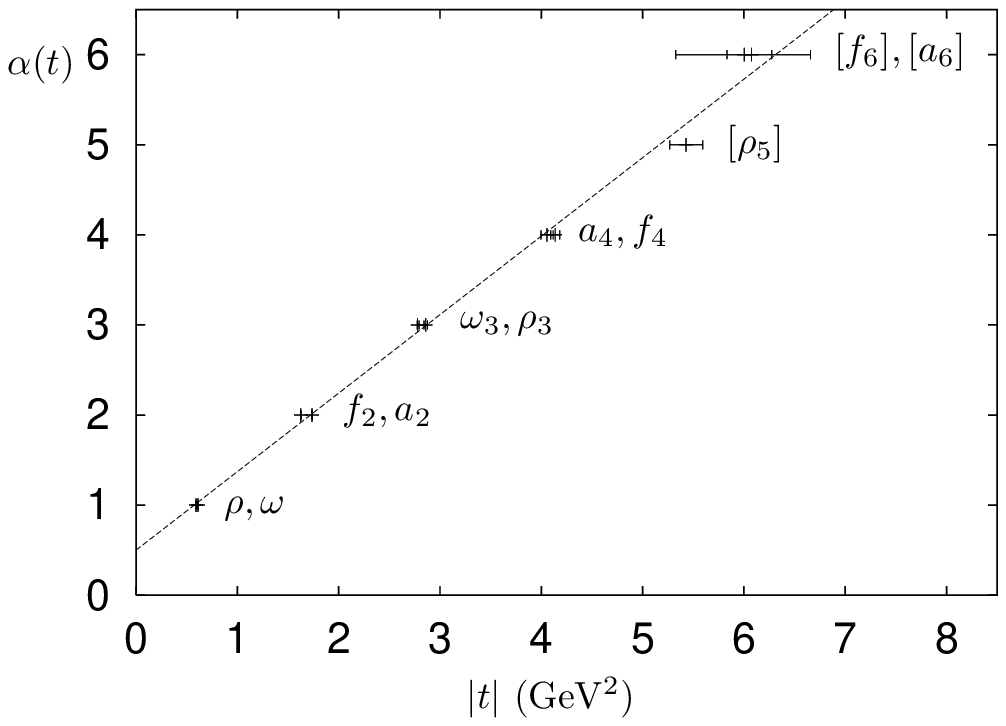}
\end{center}
\vskip -14truemm
\caption{Four degenerate Regge trajectories: particle spins plotted against
their squared masses $t$. The particles in square brackets are listed in the
data tables\cite{PDG00}, but there is some doubt about them. The straight
line is $\alpha(t)=0.5+0.9t$.}
\label{TRAJECTORY}
\begin{center}
\epsfxsize=0.4\textwidth\epsfbox[50 50 400 300]{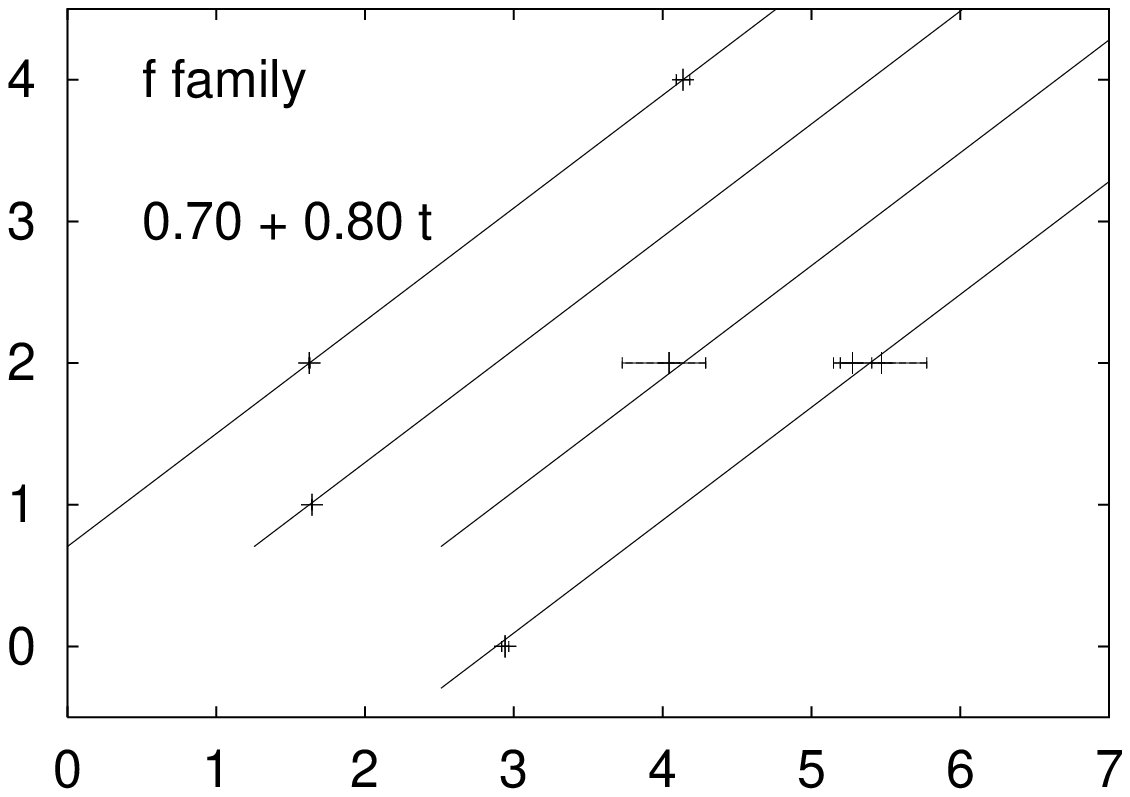}
\end{center}
\vskip -14truemm
\caption{The $f_2$ family with daughter families. Only those states
listed as confirmed in the data tables\cite{PDG00} are shown.}
\label{DAUGHTERS}
\vskip -3truemm
\end{figure}

Data for the $\pi^+p$ and $\pi^-p$ total cross-sections, with the corresponding
fits, are shown in figure~\ref{SIGTOT2}. They give just a hint that maybe a
more rapid rise than $s^{0.08}$ is indeed appropriate. It is also
interesting that the soft-pomeron coefficient is close to 2/3 that
for the $pp$ and $\bar pp$ case; that is, the soft pomeron seems to
couple to single quarks in a hadron, rather than to the whole hadron.
This is known as the additive-quark rule.

The differences between the $pp$ and $\bar pp$ cross sections, and between
the $\pi^+p$ and $\pi^-p$, are well described by a power close to
$s^{-{1\over 2}}$. This is caused predominantly by the exchange of the
$\omega$ family, and it may be seen in figure \ref{TRAJECTORY} that
the intercept of that Regge trajectory is close to $\half$.
This figure also demonstrates the well-known fact that Regge trajectories
are almost straight. We know that they cannot be exactly straight:
$\alpha_{\rho}(t)$ has a branch point at $t=4m_{\pi}^2,\, 16m_{\pi}^2,\dots$,
while $\alpha_{\omega}(t)$ has a branch point at $t=9m_{\pi}^2, \,
25m_{\pi}^2,\dots$. The surprise is how straight they are. The figure shows
also that four trajectories approximately coincide. However, this exchange
degeneracy is not exact. This is seen in figure \ref{DAUGHTERS}, which shows
that the $f_2$ trajectory has an intercept closer to 0.7 than 0.5$\,.$
This figure also shows striking confirmation for the existence of daughter
trajectories, that is trajectories spaced at integer steps below the
parent. At this meeting, Pacannoni, and more particularly Dainton,
have stressed that we might expect that the pomeron trajectory too
has daughters.
 
\section{Elastic scattering}

The single-pomeron-exchange contribution to $pp$ or $\bar pp$
elastic scattering is\cite{DL86} 
\be
{d\sigma\over dt}=C[F_1(t)]^4(\alpha' s)^{2\alpha(t)-2}
\label{elastic}
\ee
where the constant $C$ is fixed from the magnitude of the pomeron-%
exchange contribution to the total cross sections and $F_1(t)$ is
the proton's elastic form factor, measured in elastic $ep$ scattering.
It appears to be consistent with experiment to assume that the
pomeron trajectory too is straight:
\be
\alpha(t)=1+\e +\alpha' t ~~~~~~~~~~\e\approx 0.08
\label{pomtraj}
\ee
\begin{figure}
\begin{center}
\epsfxsize=0.45\textwidth\epsfbox[55 550 360 765]{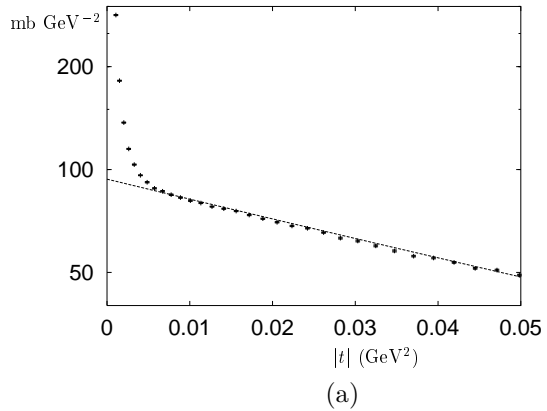}
\vskip 0truemm
\centerline{$\phantom{XXXXXX}$(a)}
\vskip 3truemm
\epsfxsize=0.45\textwidth\epsfbox[55 550 360 765]{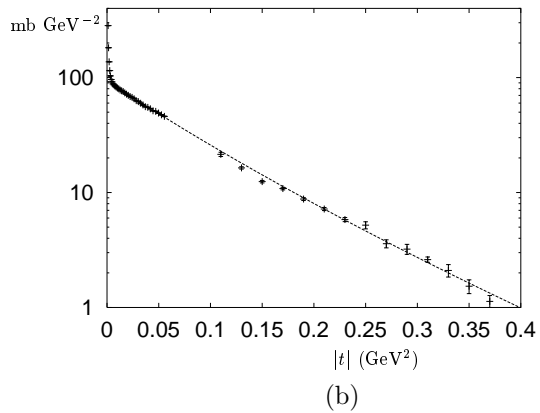}
\vskip 0truemm
\centerline{$\phantom{XXXXXX}$(b)}
\end{center}
\vskip -10truemm
\caption{$pp$ elastic scattering data at $\surd s=53$ GeV
from the CERN ISR\cite{Amo85,Bre84} with (a) the fit that determines
the value of $\alpha'_{\P}$ and (b) the fit extended to larger values
of $|t|$}
\label{ALPHAPRIME}
\end{figure}
The value of $\alpha'$ is fixed at 0.25 GeV$^{-2}$
from the data at one energy and very small
$t$: see figure \ref{ALPHAPRIME}a. The form (\ref{elastic}) then fits
the data at that energy out to larger values of $t$, as is seen in figure
\ref{ALPHAPRIME}b. This confirms that the effect of the proton wave function
is correctly taken into account with $F_1(t)$.
This again is a reflection of the apparent
fact that the pomeron couples to single quarks, like the photon\cite{LP71}.
As $F_1(t)$ is raised to the
fourth power in (\ref{elastic}), the fit is sensitive to the choice of
form factor, and it is something of a surprise that the electromagnetic form
factor, which corresponds to $C=-1$ exchange, should also be applicable
for $C=+1$ pomeron exchange. 

\begin{figure}[t]
\begin{center}
\epsfxsize=0.45\textwidth\epsfbox[55 550 360 765]{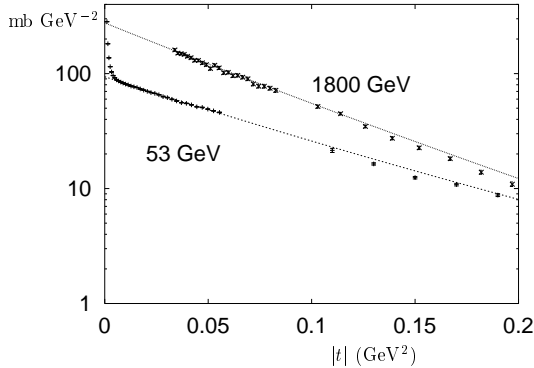}
\end{center}
\vskip -10truemm
\caption{The $\sqrt{s}= 53$ GeV data from figure \ref{ALPHAPRIME} plotted
together
with data\cite{Amo90} at $\sqrt{s}=1800$ GeV, together with (\ref{elastic})
corresponding to $\alpha'_{\P}=0.25$ in (\ref{pomtraj})}
\label{SHRINKAGE}
\vskip -5truemm
\end{figure}

Once $\alpha'$ is fixed, the form (\ref{elastic}) fits well at all energies.
In particular, it correctly predicts the shrinkage of the forward peak
in the differential cross section: see figure \ref{SHRINKAGE}. 

\begin{figure}[t]
\begin{center}
\epsfxsize=0.4\textwidth\epsfbox[45 564 340 770]{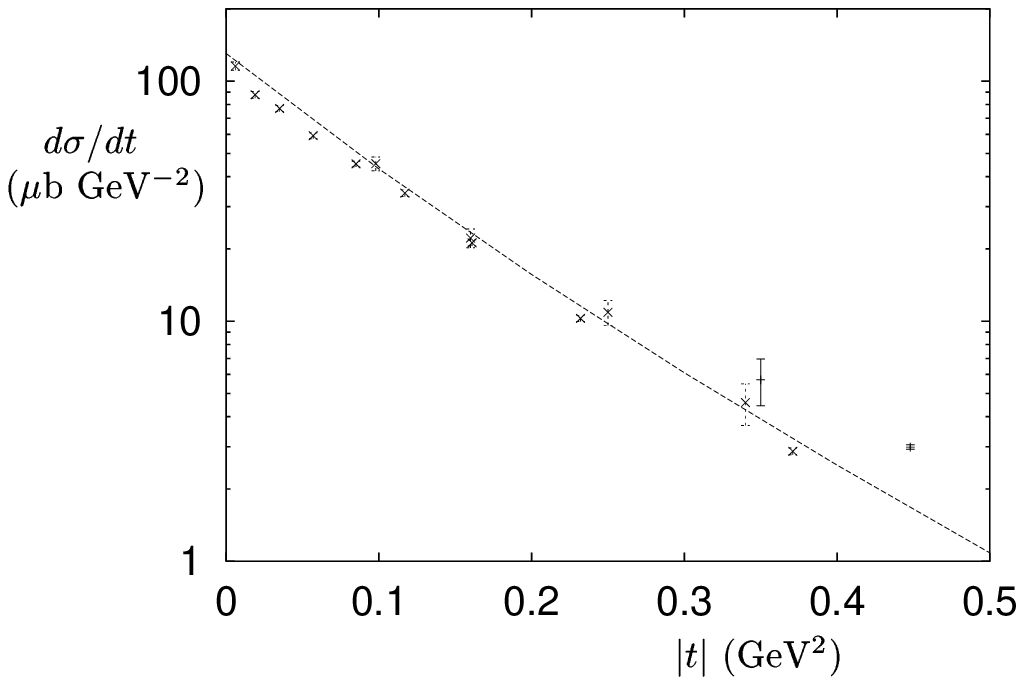}
\end{center}
\vskip -13truemm
\caption{ZEUS data\cite{Zeus98}
for $\gamma p \to \rho p$ with the pomeron-exchange
prediction. The lower $t$ data are at $\sqrt{s} = 71.7$ GeV and the higher-$t$
data at 94 GeV.}
\label{ZEUSRHO}
\vskip 2truemm
\begin{center}
\epsfxsize=0.4\textwidth\epsfbox[70 550 355 760]{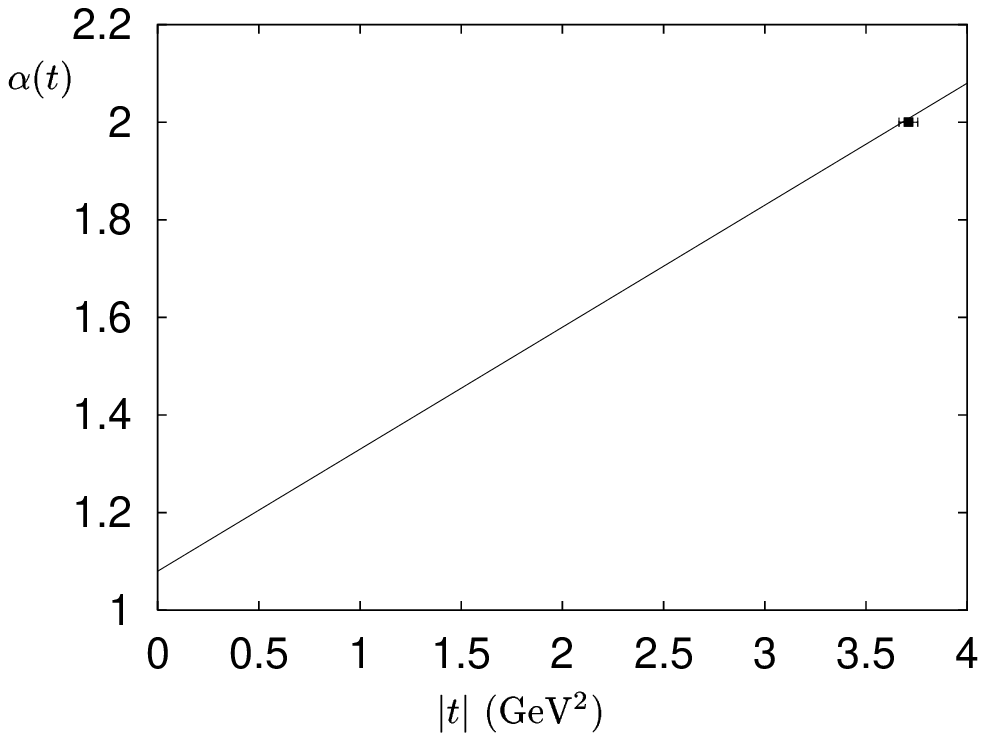}
\end{center}
\vskip -13truemm
\caption{$2^{++}$ glueball candidate\cite{WA94}, with the line
$\alpha(t)=1.08+0.25t$}
\label{GLUEBALL}
\vskip -3truemm
\end{figure}

A similar form,
with appropriate changes of form factor, also fits other elastic scattering
processes, for example\cite{DL86} $pd\to pd$. A further success is
the process $\gamma p\to\rho p$, for which one obtains\cite{DL00} a
zero-parameter fit by introducing vector meson dominance. The result is
compared with ZEUS data\cite{Zeus98} in figure \ref{ZEUSRHO}.

\section{Controversy: how large is the screening?}

Nicolescu told us at this meeting that what we know as the Froissart-Martin
bound:
\be
\sigma^{{\rm Tot}} < {\pi\over m_{\pi}^2}\log ^2s
\label{froissart} 
\ee
was first proposed by Heisenberg\cite{Hei52} as long ago as 1952. The bound
is obviously in conflict with a fixed power behaviour, so the power
$s^{0.08}$ can only be an effective power, which must reduce as $s$ increases.
There are two divergent views about this. Donnachie and I believe that
the ``bare'' pomeron yields a power just a little greater than 0.08,
and that at $t=0$ there is, at present energies,
a relatively small negative screening
contribution from the exchange of two pomerons.
Almost everybody else disagrees: the double exchange is large and the
bare-pomeron power is significantly greater than 0.08. For example, at
this meeting Kaidalov proposed the value 0.25.

As was shown by Mandelstam\cite{Man63},
the coupling of two pomerons to a hadron is not through the same quark.
Thus the notion that the screening is large at $t=0$
does not accord naturally
with the additive-quark rule, and with what we have seen is the simplest
explanation of the elastic scattering data.  
Also, there is some direct evidence\cite{Smi85} that, in the diffractive
process $pp\to\Lambda\phi Kp$, the pomeron does couple to a single quark.
Further, a large screening contribution
would make the effective power process-dependent, and
not universally close to 0.08, as various data suggest. Similar
remarks apply to the effective slope $\alpha '$ at small $t$.
A further  piece of evidence, still to be confirmed, concerns 
the question of whether there are particles on the pomeron trajectory.
It is widely agreed that, if there are, they are  glueballs. There
is a $2^{++}$ glueball candidate\cite{WA91} at exactly the right mass
to lie on a pomeron trajectory of slope $\alpha'=0.25$ GeV$^{-2}$
and intercept close to 1.08: see figure \ref{GLUEBALL}.

\section{Dips}

\begin{figure}[t]
\vskip 5truemm
\begin{center}
\epsfxsize=0.5\textwidth\epsfbox[35 575 320 775]{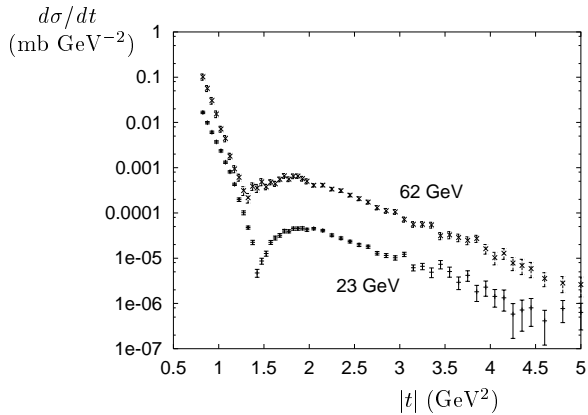}
\end{center}
\vspace{-13mm}
\caption{$pp$ elastic scattering data at larger $t$
(CHHAV collaborationx\cite{Nag79}). The 62 GeV data are multiplied by 10.}
\label{LARGEPP}
\vskip -5truemm
\end{figure}

Figure \ref{LARGEPP} shows the ISR data with the very striking  dips
in $pp$ elastic scattering. These data are reminiscent of the intensity
distributions in optical diffraction, and this has led some people to
speculate that there will be further dips at larger values of $t$.
However, although the analogy with diffraction  may be valid at small values of
$t$, and therefore the dips seen in scattering on nuclear targets
may be regarded as diffractive, for larger values of $t$ the
analogy is not all that close. In fact,
it is actually quite difficult to achieve a dip.
The reason is that the rather general requirements of analyticity
and crossing impose on an amplitude a close relationship between its
phase and its variation with energy. If we parametrise its energy 
dependence as an effective power, 
$T(s,t)\sim s^{\alpha_{\hbox{\fiverm eff}}(t)}$, then, depending on the
C-parity of the exchange responsible for this, the phase of the
$pp$ amplitude is 
\beqa
-e^{-{1\over 2}i\pi\alpha_{\hbox{\fiverm eff}}(t)}~~~~~~~~~~~&&C=+1\nn\\
-ie^{-{1\over 2}i\pi\alpha_{\hbox{\fiverm eff}}(t)}~~~~~~~~~~~&&C=-1
\label{phase}
\eeqa
As is seen in figure \ref{LARGEPP}, the dip has steep sides and moves
inwards as $s$ increases. Hence if we fix $t$ at some value in the dip
region, the amplitude varies rapidly with energy. So models in which
the amplitude is taken to be pure-imaginary for all $t$ are not 
consistent with (\ref{phase}). 

It is natural to try to achieve a dip through interference between
single-pomeron and two-pomeron exchange.  Although we cannot calculate
the strength of two-pomeron exchange, we do know that it yields a cut
in the complex angular momentum plane. If the pomeron trajectory is
(\ref{pomtraj}). the cut trajectory is
\be
\alpha_{\P\P}(t)=1+2\e+\half\alpha' t
\label{cuttraj}
\ee
With $\e=0.08$ and $\alpha'=0.25$, the phases associated with the single
and double exchanges are
\beqa
-e^{-{1\over 2}i\pi\alpha_{{\P}}(t)}&=&-0.4+0.9i\nn\\
-e^{-{1\over 2}i\pi\alpha_{{\P\P}}(t)}&=&-0.02+1.0i
\label{phases}
\eeqa
So if there is destructive interference between the imaginary parts
of the two contributions, we still need something else to help cancel
their real parts. Donnachie and I suggested\cite{DL83a} that the extra
term was 3-gluon exchange, and therefore predicted that the dip would
not be present in $\bar pp$ elastic scattering, because 3-gluon exchange
is $C=-1$ and therefore contributes with opposite signs to $pp$ and
$\bar pp$. This prediction was subsequently confirmed\cite{Bre85},
and it is now fairly generally accepted \cite{GLN84,PF00} that to the left
of the dip the dominant exchange is $C=+1$, while to the right it is $C=-1$.

\begin{figure}[t]
\hbox{\hskip -5truemm\epsfxsize=0.5\textwidth\epsfbox[54 484 436 760]{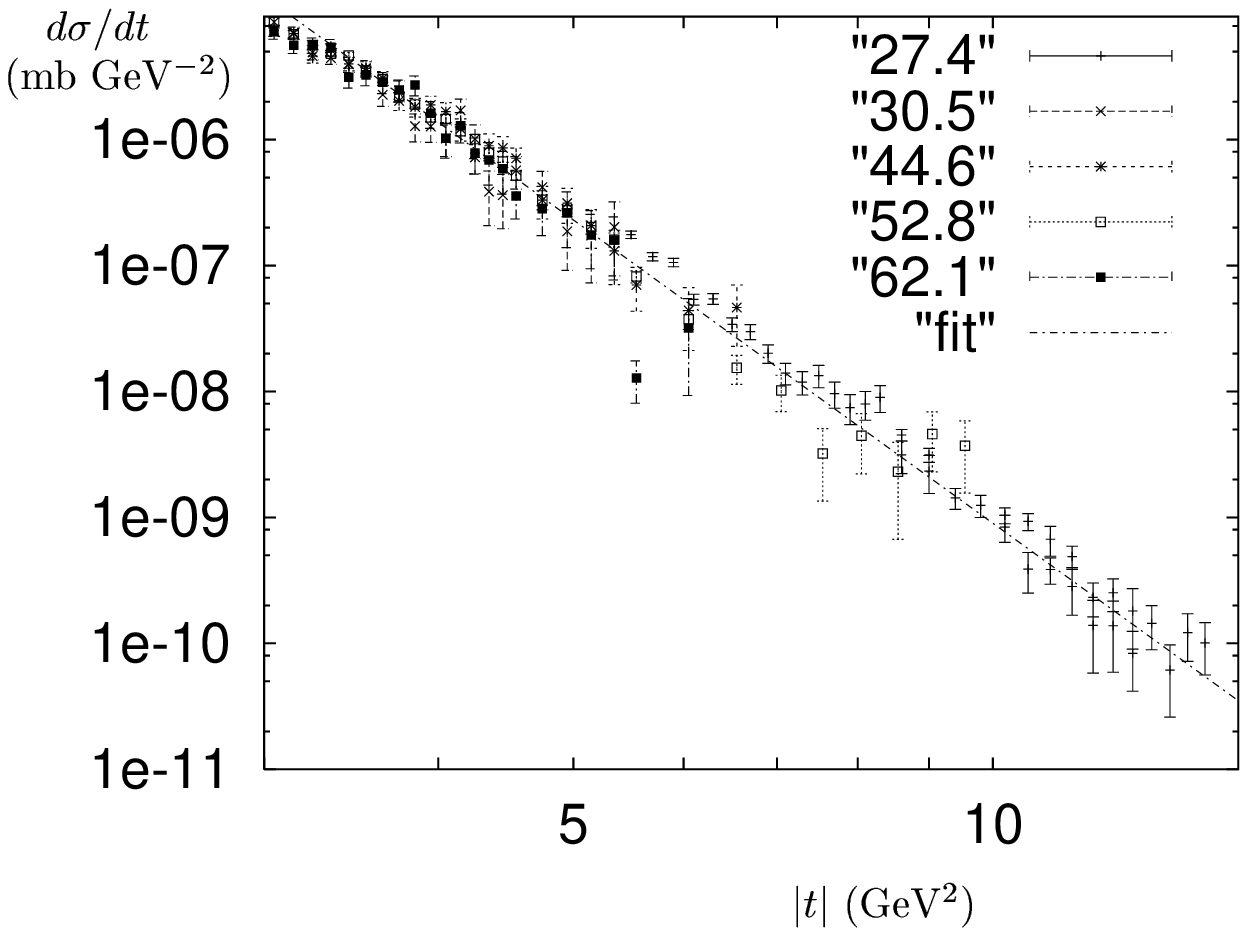}}
\vskip -9truemm
\caption{$pp$ elastic scattering data\cite{Nag79,Fai81}
at the largest available $t$; the line is $0.09\,t^{-8}$}
\label{LARGE-T}
\vskip 8truemm
\begin{center}
\epsfxsize=0.35\textwidth\epsfbox[0 0 256 182]{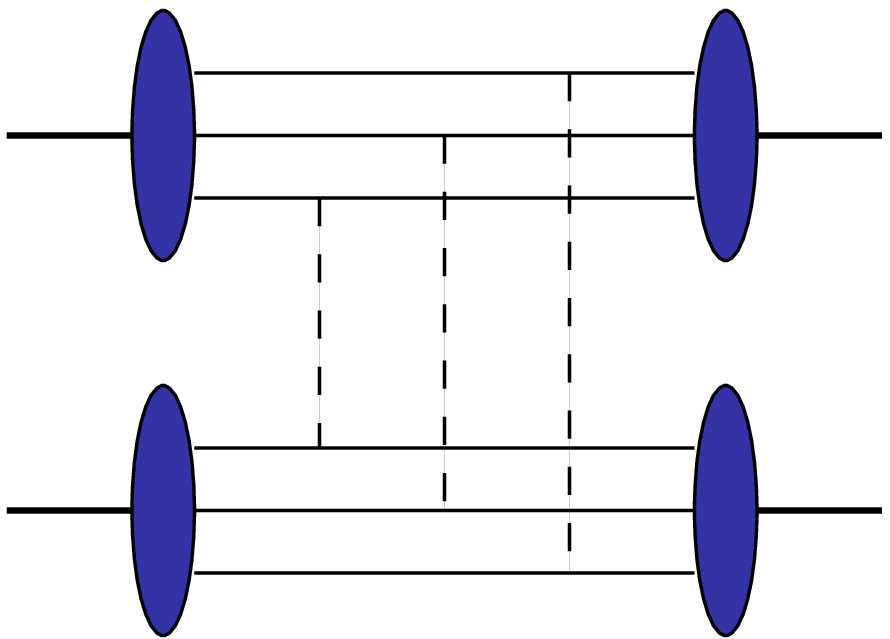}
\end{center}
\vskip -6truemm
\caption{3-gluon exchange}
\label{GGG}
\vskip -4truemm
\end{figure}

Figure \ref{LARGE-T} shows $pp$ data at $t$-values beyond the dip.
As is seen, for $\sqrt s>27$ GeV the data are almost energy-independent
and behave as $t^{-8}$. This is what is what is calculated\cite{DL79}
from perturbative triple-gluon exchange at large $|t|$ to lowest order
in the QCD coupling (figure \ref{GGG}).

It is interesting to ask whether the energy-independence at large $t$
survives at higher energies. In particular, if there really is a 
BFKL pomeron, replacing the gluons in figure \ref{GGG} with it might
give a dramatic {\it rise} with increasing energy.
Again a question for the LHC.
\eject

\section{Diffraction dissociation}

\begin{figure}[t]
\begin{center}
\epsfxsize=0.4\textwidth\epsfbox[50 50 390 290]{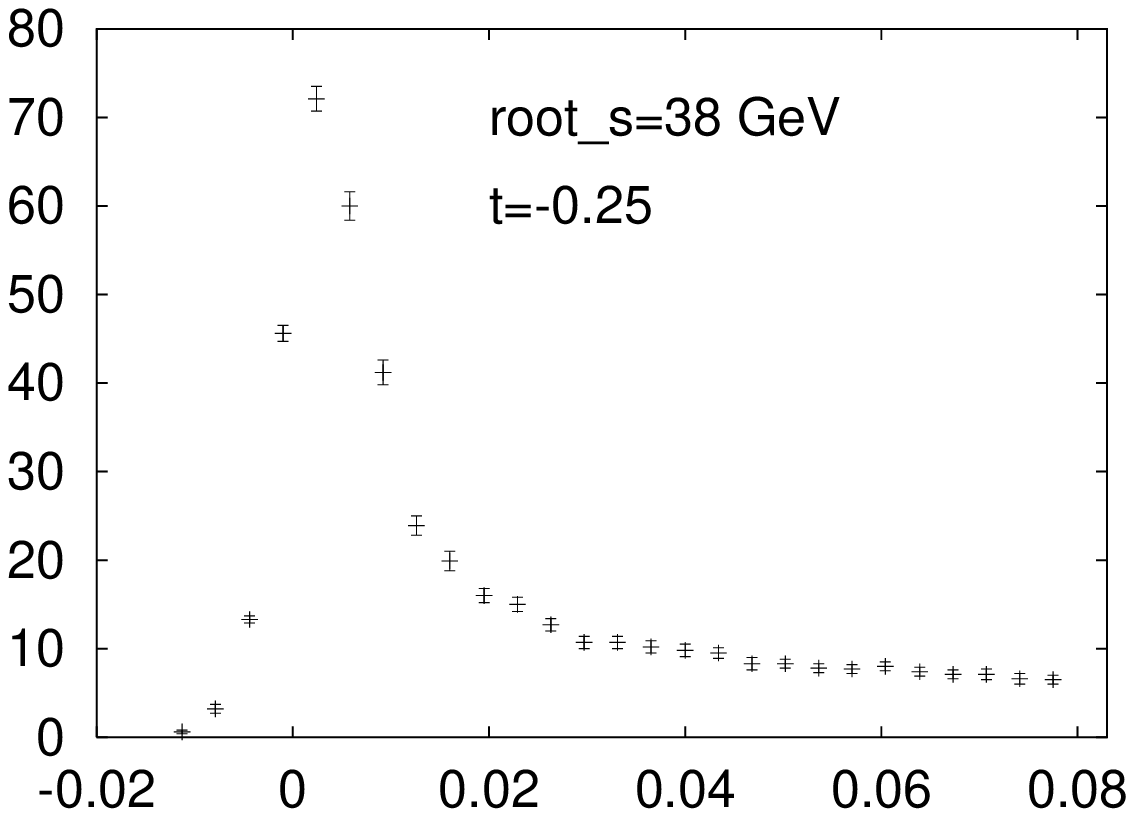}
\end{center}
\vskip -13truemm
\caption{CHLM data\cite{Arm82}: $d^2\sigma/dtd\xi$ plotted against
$\xi=M_X^2/s$}
\label{CHLM}
\vskip 8truemm
\begin{center}
\epsfxsize=0.25\textwidth\epsfbox[0 0 435 380]{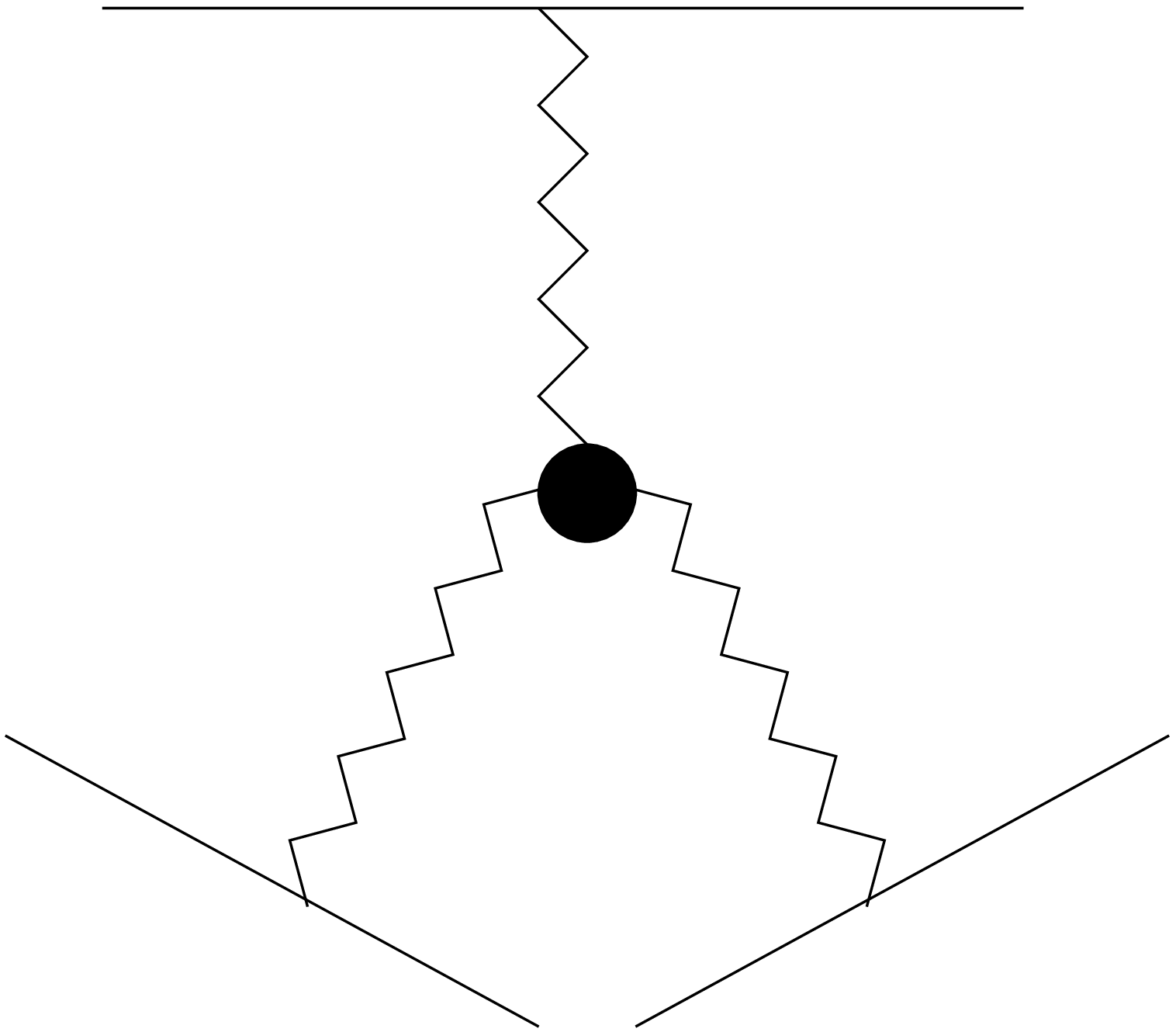}
\end{center}
\vskip -11truemm
\caption{Triple-reggeon mechanism}
\label{TRIPLE}
\vskip -5truemm
\end{figure}

Figure \ref{CHLM} shows data for diffraction dissociation,
$pp\to pX$, where the final-state proton has lost only a small fraction
$\xi$ of its initial momentum. Experimentalists often quote the value
of the ``total'' diffractive  cross section, which is $d^2\sigma/dtd\xi$
integrated over some range of $t$ and of $\xi=M_X^2/s$. It is evident
from the figure that there is a resolution problem: $\xi$ cannot
really be negative. There is uncertainty about the correction for this.
Also, most of the cross section comes from small $M_X$ (and small $t$),
and so the answer is sensitive to the choice of lower limit in the
integration.  

Apart from these experimental uncertainties, there are serious theoretical
ones. It is often assumed that the mechanism is that of figure \ref{TRIPLE},
the triple-pomeron diagram. However, unless $M_X$ is large, say greater
than 50 GeV, there is also a significant contribution from the diagram
where the upper pomeron is replaced with a nonleading exchange. And if
$M_X$ is too small, less than a few GeV, the mechanism is not applicable
at all. Again, unless $\xi$ is very small, there are significant contributions
from either of both of the lower pomerons being relaced with a nonleading
exchange ($f_2,\omega,\pi\dots$). Fits to data are very model-dependent,
but they always find that these contaminations from nonleading exchanges
are large\cite{RR74,DL84}. An example is the H1 fit to their data\cite{H197}
for real-photon-induced diffraction dissociation, figure 14.
\begin{figure}[t]
\begin{center}
\vskip -6truemm
\epsfxsize=0.45\textwidth
\epsfbox[43 460 490 770]{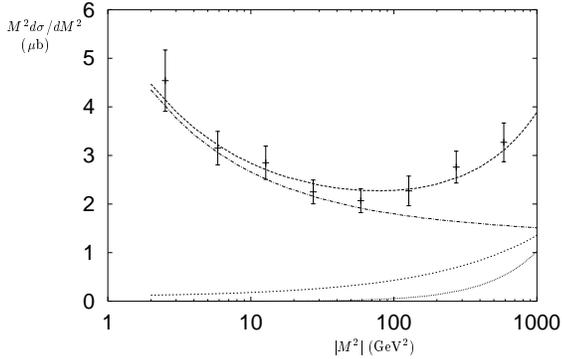}
\vskip -11truemm
\caption{The diffraction dissociation cross section $M_X^2d\sigma/dM_X^2$
in photoproduction at $W = 187$ GeV\cite{H197}
The lines, from the bottom upwards, represent fits to the reggeon-exchange
contribution, its interference with pomeron exchange, the pomeron exchange,
and the total.}
\end{center}
\label{XXXXX}
\vskip -9truemm
\end{figure}

At this meeting, Goulianos has suggested that there are problems
with the conventional Regge interpretation of the data for the total
diffractive cross section. For the reasons I have given, I do not
believe that there are yet grounds to worry. We need more detailed data,
which unfortunately have not been provided yet by the Tevatron experiments.
There are, as yet, do data at higher energies as good as the ISR
data, an example of which is shown in figure \ref{CHLM}.

\section{Controversy: how many pomerons?}

I turn now to hard processes. The data show clearly that the soft
pomeron is not enough, and something else is needed. We call this
the ``hard pomeron'', though there are different views on what this
is. However, even without this question, there is a more basic
one: is it that the soft pomeron becomes progressively harder as the
scale $Q^2$ of a reaction increases, or is the hard pomeron a separate
object from the soft one?

Specifically, for the case of the small-$x$ behaviour
of the structure function $F_2(x,Q^2)$,
the first of these two alternatives is
\be
F_2(x,Q^2)=f_0(Q^2)x^{-\e _0}+f_1(Q^2)x^{-\e _1}
\label{alt1}
\ee
with  both $\e _0$ and $\e _1$ fixed. Here $\e _1$ is the soft-pomeron
power, $\e _1\approx 0.08$, and the data need\cite{DL98} $\e _0$ to
be a little greater than 0.4$\,$. As $Q^2$ increases, the ratio
$f_0(Q^2)/f_1(Q^2)$ increases. This is the alternative that Donnachie
and I prefer, because we take seriously the prejudice developed
40 years or so ago that the positions of singularities in the complex angular
momentum plane should not vary with $Q^2$.

Almost everybody else adopts the alternative approach, that at small $x$
\be
F_2(x,Q^2)={\cal F}(Q^2)x^{-\e (Q^2)}
\label{alt2}
\ee
where $\e (Q^2) =\e _1$ at $Q^2=0$ and rises to $\e _0$, or even larger,
at large $Q^2$.
This rise in the value of $\e (Q^2)$ is supposed to
be caused by perturbative evolution; however, there is a growing
realisation that we do not understand the theory of perturbative evolution
at small $x$. Martin gave some hint of this in his talk.

\begin{figure}[t]
\vskip -7truemm
\begin{center}
\epsfxsize=0.45\textwidth\epsfbox[50 50 410 300]{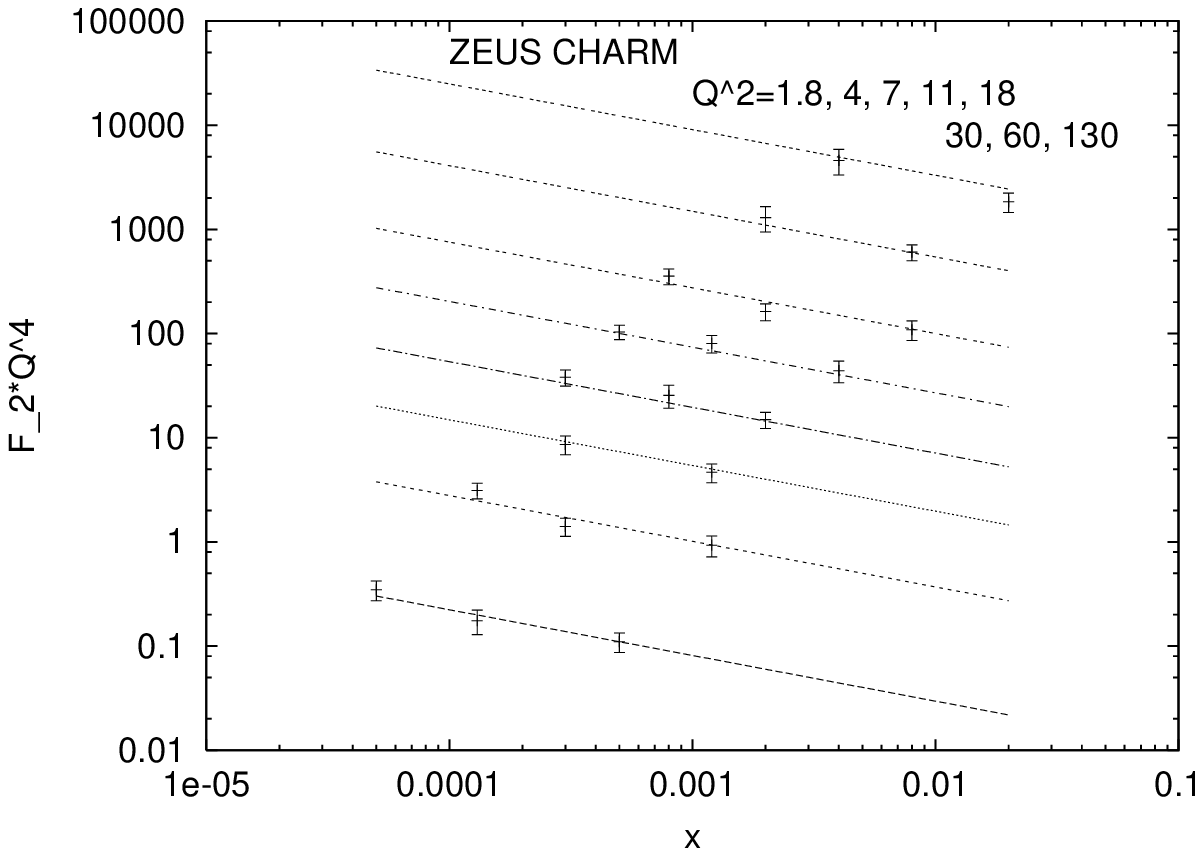}
\end{center}
\vskip -11truemm
\caption{Data\cite{Z00} for the $Q^4F_2^c(x,Q^2)$, fitted\cite{DL00}
to $f_0(Q^2)x^{-\e _0}$ with $\e _0=0.44$}
\label{CHARM}
\vskip 3truemm
\begin{center}
\epsfxsize=0.45\textwidth\epsfbox[50 50 410 300]{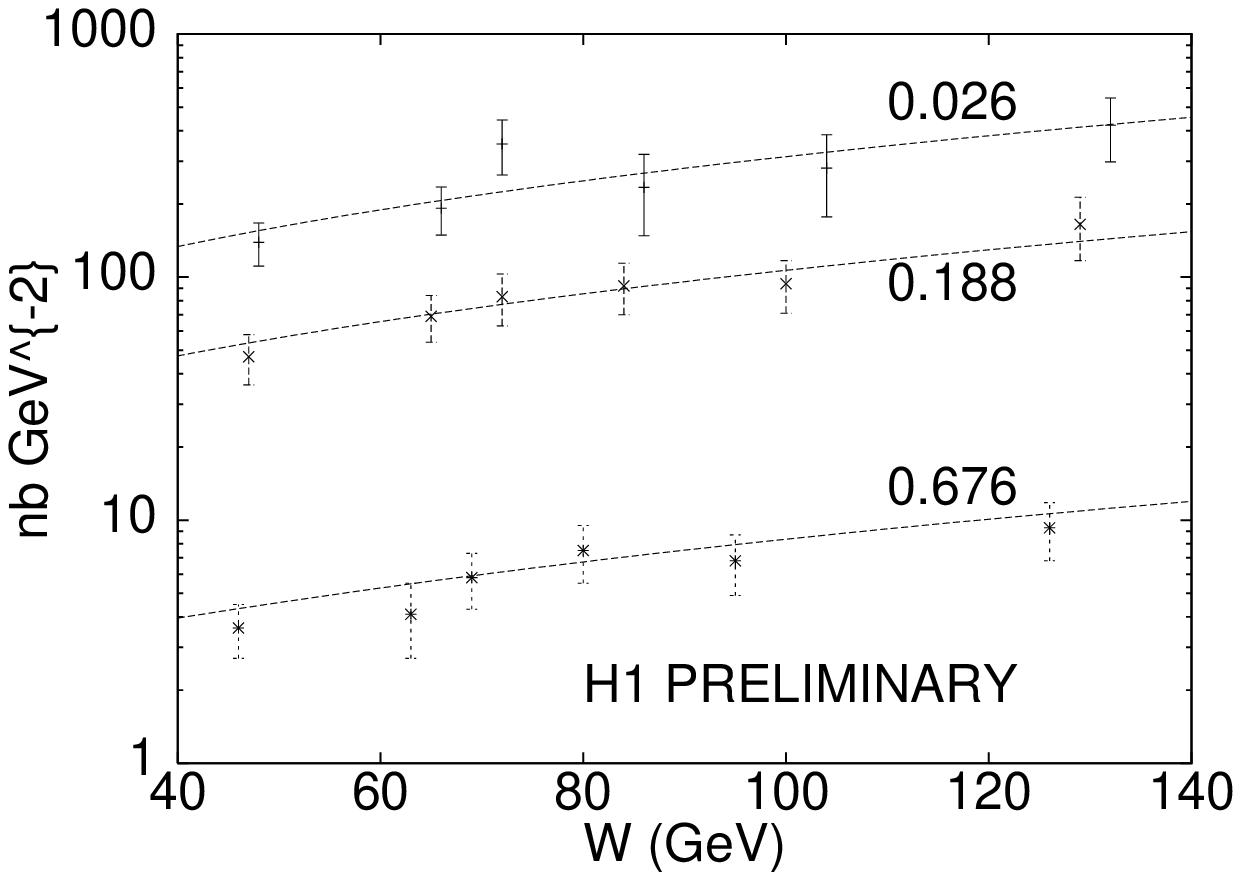}
\end{center}
\vskip -11truemm
\caption{The differential cross section for $\gamma p\to J/\psi\, p$}
\label{JPSI}
\vskip -5truemm
\end{figure}

The ZEUS data for the charm component of $F_2$, that is the part $F_2^c$ of
$F_2$ for which the virtual photon is supposed to have been absorbed
by a charmed quark, seems to provide striking confirmation\cite{DL99b} of the
fixed-power approach: see figure \ref{CHARM}. However, these data need
to be treated with some caution, because to extract them a large
extrapolation in $p_T$ is needed. Nevertheless, the same approach
gives an excellent description\cite{DL00} of the process
$\gamma p\to J/\psi\, p$, both the total cross section and the $t$ dependence. 
The fit shown in figure \ref{JPSI} corresponds to the amplitude
\be
T(s,t)=i\sum_{i=0,1}C_iF_1(t)s^{e_i(t)}e^{-{1\over 2}i\pi e_i(t)}
\label{jpsi1}
\ee
with
\beqa
e_0(t)&=&0.44+0.1t\nn\\
e_1(t)&=&0.08+0.25t
\label{jpsi2}
\eeqa
$F_1(t)$ is again the Dirac form factor of the proton target,
and  $C_0$ and $C_1$ are both independent of $s$ and $t$. 
It is interesting that these data
lead to a hard-pomeron trajectory slope that is somewhat smaller than that
of the soft pomeron.

For $\gamma p\to J/\psi\, p$ the ratio of $C_0$ to $C_1$ needs to
be about ${1\over 10}$. A similar fit to $\gamma p\to \rho p$
needs this ratio to be much smaller, mainly because the soft-pomeron
term $f_1$ is much larger. If we extend the fit to
$\gamma^* p\to \rho p$, the ratio must grow with $Q^2$.

\begin{figure}[t]
\begin{center}
\epsfxsize=0.45\textwidth\epsfbox[50 50 750 550]{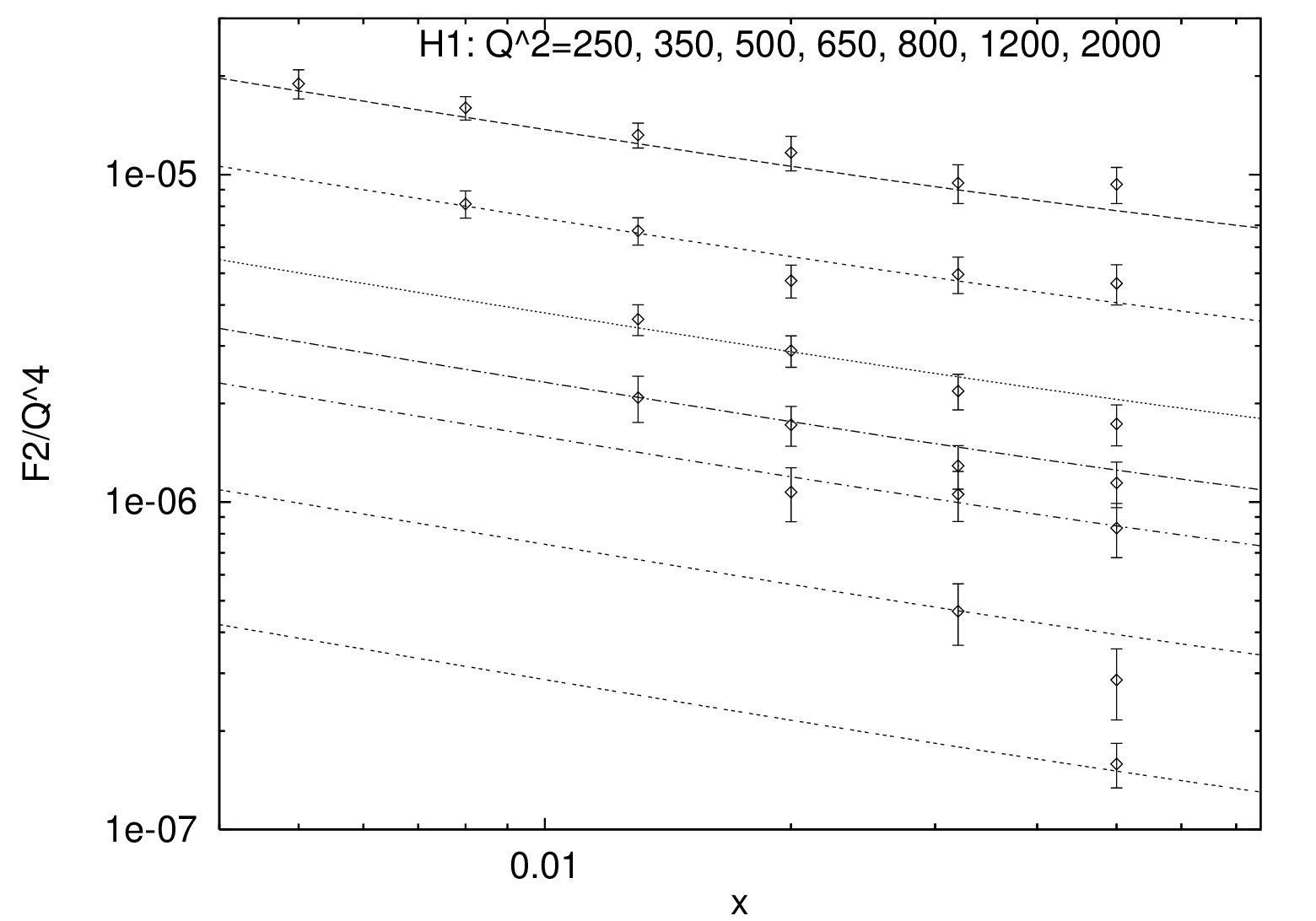}
\end{center}
\vskip -12truemm
\caption{Large-$Q^2$ data for $F_2(x,Q^2)$ with \hbox{$x<0.07$}}
\label{H1LARGEST}
\begin{center}
\epsfxsize=0.45\textwidth\epsfbox[50 50 410 300]{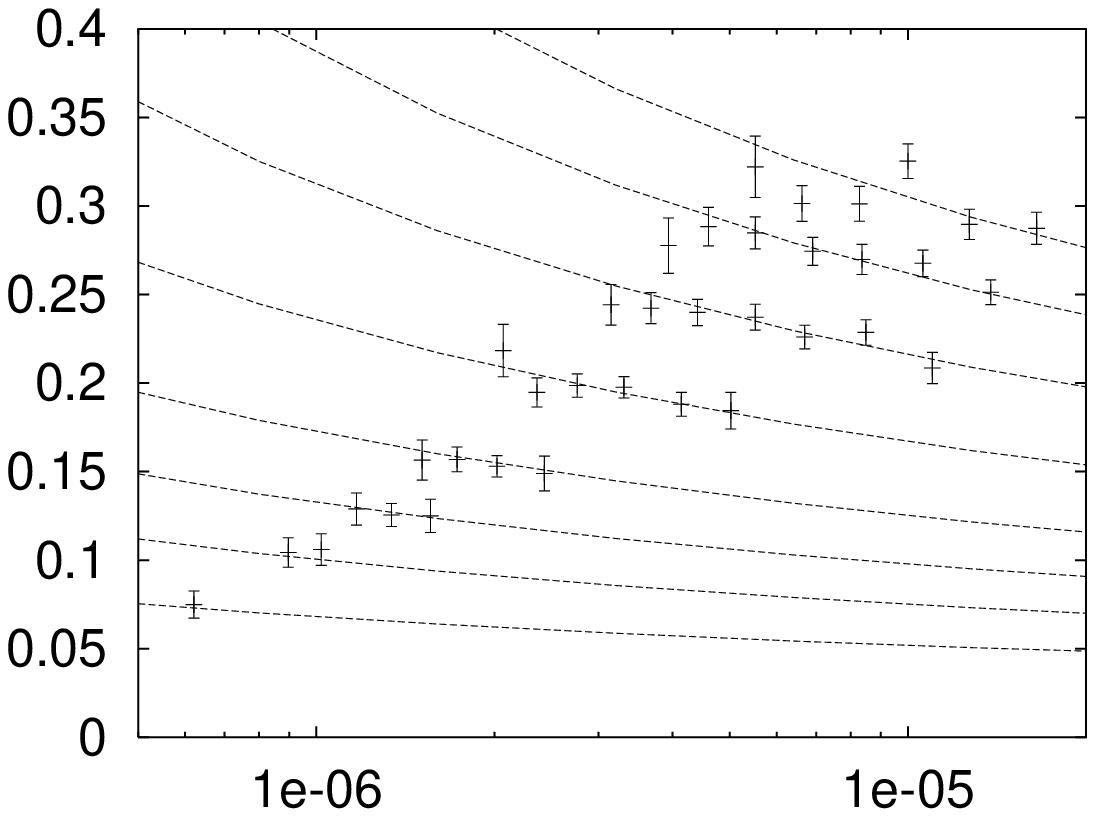}
\end{center}
\vskip -12truemm
\caption{Small-$Q^2$ data, ranging from \hbox{$Q^2=0.045$} (lower points)
to 0.3 GeV$^2$ (upper points)}
\label{ZEUSSMALLEST}
\end{figure}

\begin{figure}[t]
\vskip -2truemm
\begin{center}
\epsfxsize=0.45\textwidth\epsfbox[50 50 410 300]{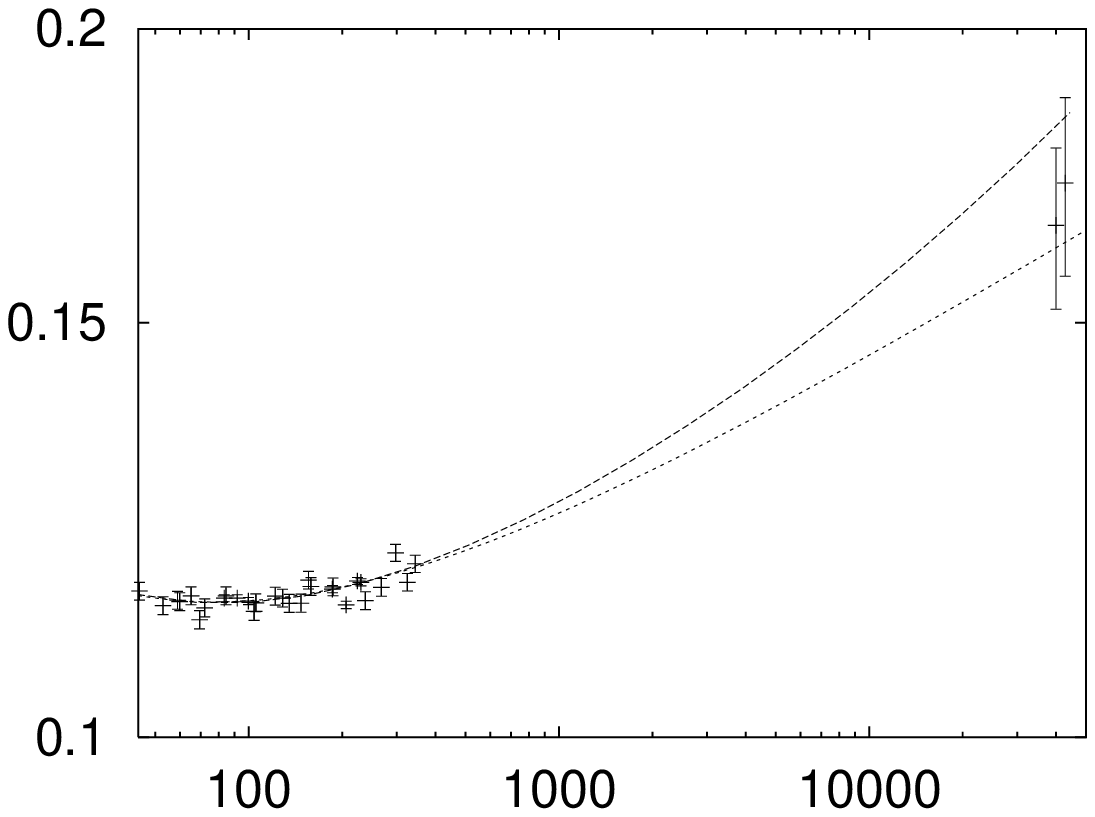}
\end{center}
\vskip -12truemm
\caption{Data for $\sigma^{\gamma p}$. The upper line is an extrapolation
of the data of figure \ref{ZEUSSMALLEST} to $Q^2=0$.}
\label{ROUTPUT}
\vskip 6truemm
\begin{center}
\epsfxsize=0.42\textwidth
\epsfbox[107 575 335 760]{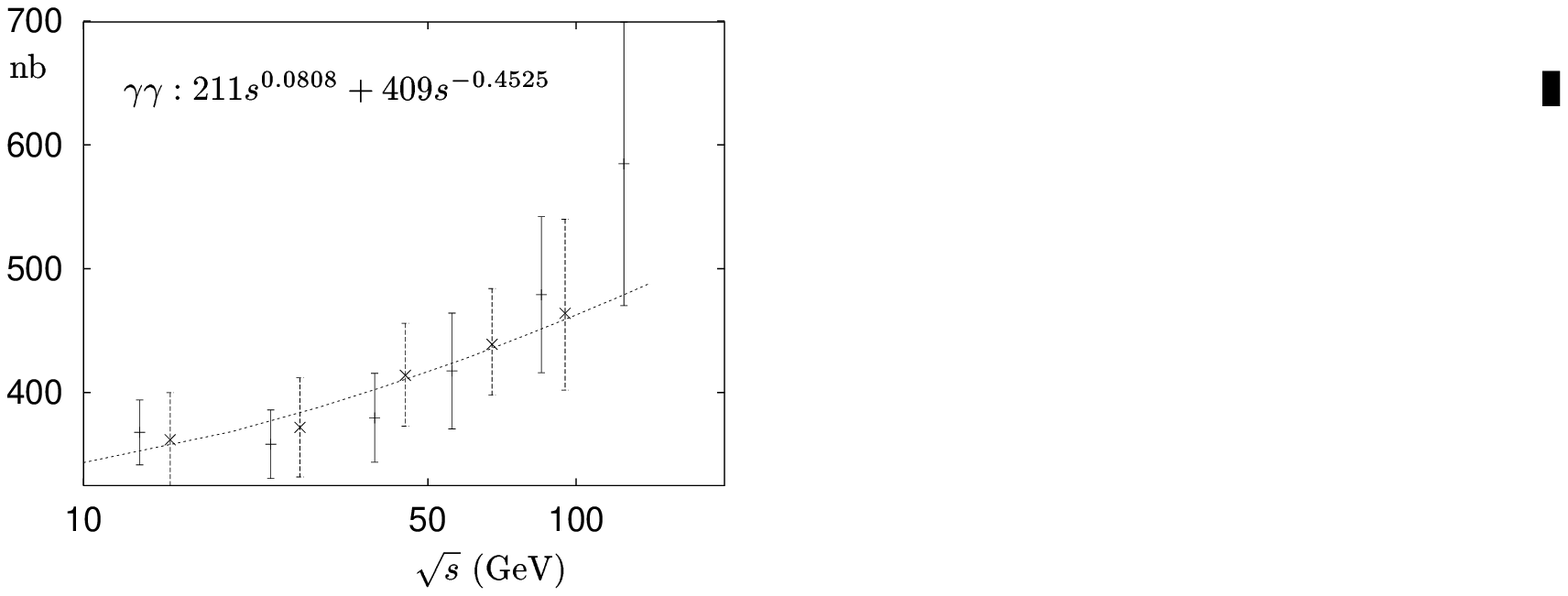}
\end{center}
\vskip -10truemm
\caption{$\gamma \gamma$ total cross section with the
fit obtained from factorisation}
\vskip -7truemm
\label{GG}
\end{figure}

The form (\ref{alt1})
provides\cite{DL98} an excellent fit to the small-$x$ data for $F_2(x,Q^2)$.
With an additional term 
$f_2(Q^2)x^{0.45}$ corresponding to $f_2$ exchange, the fit extends up to
$x=0.07$ or more, and it covers all the available range of $Q^2$,
from real photons to $Q^2=2000$ GeV$^2$. Figures \ref{H1LARGEST}
and \ref{ZEUSSMALLEST} show data at large and small $Q^2$, while
figure \ref{ROUTPUT} shows data for real photons. 

\section{Key question: is the hard pomeron present at $Q^2=0$?}

The lower curve in figure \ref{ROUTPUT} shows the 
prediction for $\sigma^{\gamma p}$ that we made\cite{DL92}
before the HERA data were available, and including only soft-pomeron
and $f_2$-exchange terms. The highest-energy data point is the new
ZEUS measurement, which now agrees well with the H1 point. The upper
line is the extrapolation to $Q^2=0$ of the small-$Q^2$ ZEUS data
of figure \ref{ZEUSSMALLEST}. This extrapolation includes a hard-pomeron
component. Because the data in figure \ref{ZEUSSMALLEST} extend down
to very small $Q^2$, this extrapolation would be rather reliable, were
it not for the fact that the data in figure~\ref{ZEUSSMALLEST} do not
show the systematic errors, which are as large as $\pm 10$\% at the lowest
$Q^2$. So we cannot be sure whether a hard pomeron component is present
at $Q^2=0$ or not, though there is more than a hint that it may be.

There is a similar uncertainty with the interpretation of the data
for $\sigma ^{\gamma\gamma}$ from LEP. De Roeck has told us at this
meeting that the L3 collaboration favours the presence of a hard-pomeron
component at $Q^2=0$, while OPAL does not. Figure \ref{GG} shows the fit
obtained by applying factorisation to the $\sigma^{\gamma p}$ and
$\sigma^{pp}$ curves in figures (\ref{ROUTPUT}) and (\ref{SIGTOT1}),
with no hard-pomeron component. The highest-energy data point from L3
possibly does call for such a component to be added.

This is an important question. Is the hard pomeron already present at
$Q^2=0$, or is rather generated by perturbative evolution?
If it is there already at $Q^2=0$, since it is not seen in the
$\bar pp$ total cross section presumably it arises because the photon
has a pointlike component. But occasionally the quarks in a proton
are very close together, so a hard-pomeron component should also
be present in $\sigma ^{pp}$. Will it be large enough to be identified at
the LHC?

\section{Hard diffraction}

The present situation concerning hard diffraction is a confusing one.
The two HERA experiments agree that the effective pomeron intercept
in hard-diffractive events is significantly higher than 1.08,
$\alpha_{\hbox{\fiverm eff}}\approx 1.2$, and that the gluons provide
80 to 90\% of the momentum of the pomeron. Goulianos told us at this
meeting that the Tevatron experiments conclude that the gluon momentum
fraction is no more than 50\%, and that they find a dramatic
breakdown in factorisation.

There are theoretical uncertainties. First, if a diffractive event is
defined as one with a very fast proton, the theory is quite well
defined. But in practice much of the data is rather obtained by requiring
just a large rapidity gap, for which the theory is much less certain.
We have heard Khoze describe an apparently-successful calculation of gap
survival probabilities, but the theory of this is deeply uncertain.
Even apart from this difficulty, 
until it becomes possible to compare data from HERA and the Tevatron
where both triggers are on a very fast final-state proton (or antiproton)
there is a worry that we may not be comparing like with like, if what goes
down the beam pipe is not the same. Again, there are uncertainties
in extracting the gluon component of the pomeron structure function;
the HERA experiments have obtained rather different outputs using
different analyses.
Further, one should not expect
factorisation, given that $\alpha_{\hbox{\fiverm eff}}$ is not equal to
the soft-pomeron value. Whatever is the explanation for this, whether
it be BFKL, a combination of hard and soft pomerons, important screening
effects, or any other, all would agree that factorisation should break
down, Nevertheless, it is surprising that the breakdown is so dramatic.
For example, Goulianos and Santoro have told us that
the Tevatron cross section for $W$ production acompanied by
a very fast proton or antiproton is nearly an order of magnitude
smaller than expected, and diffractive dijet production has similar
problems. 

\section{Diffractive Higgs production}

\begin{figure}[t]
\begin{center}
\epsfxsize=0.2\textwidth\epsfbox{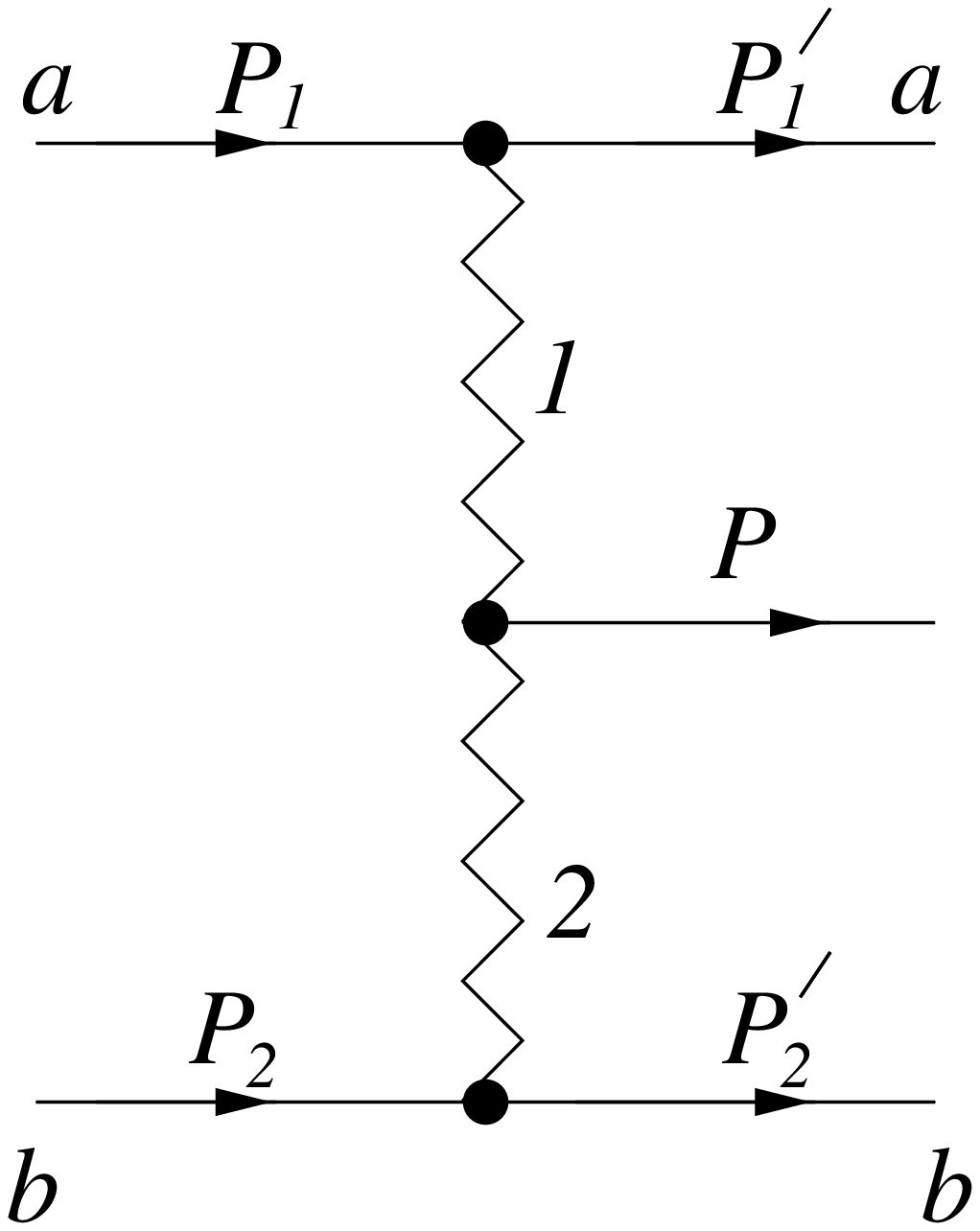}
\end{center}
\vskip -7truemm
\caption{Exclusive central Higgs production}
\label{PRODUCTION}
\vskip 4truemm
\begin{center}
\epsfxsize=0.2\textwidth\epsfbox{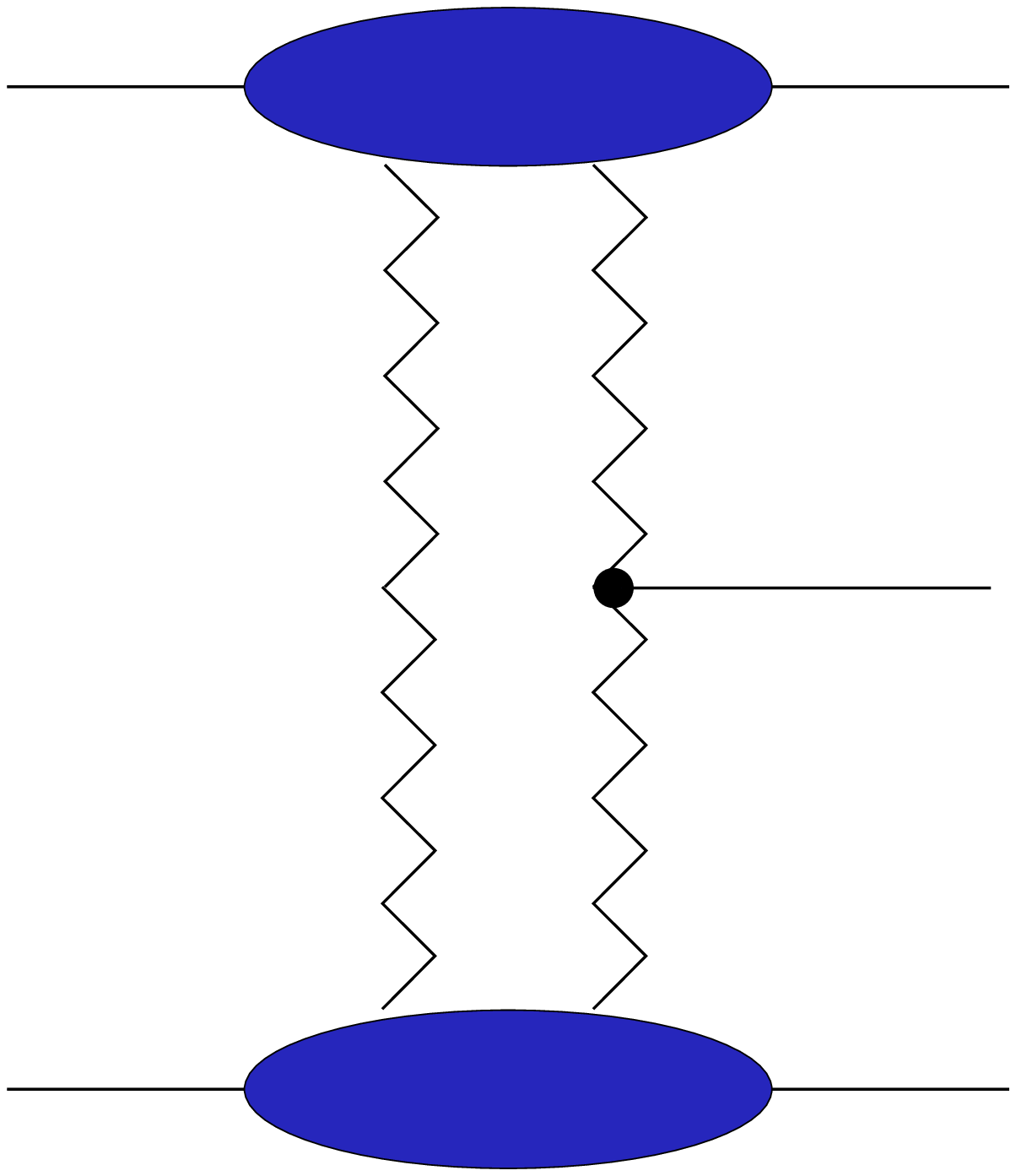}
\end{center}
\vskip -9truemm
\caption{Screening correction to figure \ref{PRODUCTION}}
\label{HIGGSCREEN}
\vskip -5truemm
\end{figure}

It was suggested 10 years ago\cite{SNS90} that diffractive Higgs production
should be feasible: see figure \ref{PRODUCTION}. This is the process
$pp\to pHp$, with both final-state protons very fast. There have been
various calculations that model the coupling of the 
two pomerons to the Higgs. Most\cite{BL91,CH96}, though not 
all\cite{GLM95,KMR00}, predict a cross section large enough for it to have been 
suggested\cite{AR00} that this
exclusive reaction  might
be the best way to discover the Higgs particle. This is because the
momenta of the two final protons  may
be measured very accurately, so that the missing mass, presumed to
be that of the Higgs, may be determined very accurately, and in consequence
the signal-to-background ratio is much enhanced. The reason for this is
as follows.  The best way to detect the Higgs is through its decay
to a heavy quark-antiquark pair. The background is direct production
of the pair, not through a Higgs. The ratio of the amplitudes for
signal compared with the background is similar\cite{BS92} to
that in the case of ordinary, nondiffractive production. However, because
of the much better mass resolution, the background is integrated over a
much smaller mass range.

The difference between those who predict a large cross section and those
who do not lies in the estimate of the screening correction to 
figure \ref{PRODUCTION}: see figure \ref{HIGGSCREEN}. If you believe,
as I do, that screening in the $pp$ total cross section is small, then
it is hard to believe that it will reduce the diffractive Higgs production
by a factor of more than 2, say.  But those who believe in large screening
believe that the suppression is an order of magnitude or more.

\section{Perturbative evolution}

\begin{figure}[t]
\begin{center}
\epsfxsize=0.4\textwidth\epsfbox[50 50 390 290]{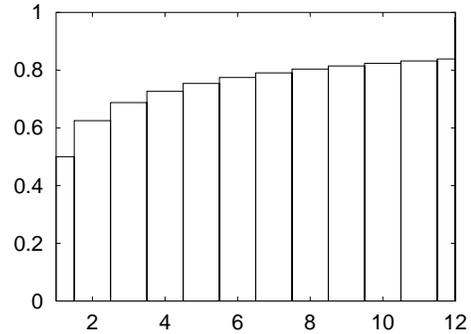}
\end{center}
\vskip -13truemm
\caption{Partial sums of the series expansion of (\ref{sqr}) at
$x=1$ plotted against the number of terms}
\label{SQR}
\end{figure}
There has been a growing realisation that we do not understand the theory
of perturbative evolution at small $x$. While there are various ideas
on the subject, there is no agreement whatever\cite{BL00}. 
The problem lies in the perturbative expansion of the DGLAP splitting
function $P(z)$, whose terms are singular at $z=0$. In lowest order,
there are terms $\alpha_S/z$, and the Mellin transform has terms 
$\alpha_S/N$. In higher orders the singularities at $z=0$ or $N=0$
become worse. For small-$x$ evolution, we need the splitting function
for small values of $z$ or $N$, so the expansion parameter is large and
the expansion is illegal.

It is not neccessarily a comfort to ignore this problem and find that
the perturbation expansion stabilises after a few terms. As a simple
example, consider the expansion of the function
\beqa
f(x)&=&1-\sqrt{1-x}\nn\\
&=&\half x+{1\over 8}x^2 +{1\over 16}x^3+
{5\over 128 x^4}+\dots
\label{sqr}
\eeqa
at $x=1$. The result is shown in figure \ref{SQR}. After 10 terms,
each additional term in the expansion contributes less than 1\%, but 
nevertheless the sum is nearly 20\% below the correct value.

So, finding that higher-order terms in a perturbation expansion are
small does not necessarily indicate that we are near to the right
answer. We should not trust the expansion for values of $x$ and $Q^2$
where it leads to a rapid variation of $F_2(x,Q^2)$. That is, while it
might possibly be correct that the hard pomeron is generated by evolution,
this cannot be trusted numerically until $Q^2$ becomes large.

The known connection\cite{J82} between DGLAP and BFKL tells us
that $P(N)$ is not singular at $N=0$, even though each term in its
expansion is singular. In fact $P(N)$ behaves something like
$f(x)$ in (\ref{SQR}) with $x$ set
equal to $\alpha_S/N$:  the partial sums become more and more
singular at $N=0$ as we add terms, but the function itself is not
singular at $N=0$. 

My guess is that $P(N)$ has no relevant singularities at all. In that
case\cite{CDL99}, a fixed power $f_0(Q^2)x^{-\e _0}$ of $x$ remains
a fixed power under evolution, and the evolution determines the behaviour
of $f_0(Q^2)$ at large $Q^2$. that is, the hard pomeron is already 
present at $Q^2=0$ and the evolution presumably enhances its relative
importance as $Q^2$ increases.

\section{The main issues}

\bu Is screening in soft processes large or small?

\bu Are the hard and soft pomeron distinct objects?

\bu Is a hard pomeron already present at $Q^2=0$? Will it be seen in
the $pp$ total cross section at the LHC?

\bu Are there glueballs on pomeron trajectories?

\bu How straight are the trajectories?

\bu Can we calculate gap survival probabilities?

\bu Do screening corrections leave $pp\to p \;{\rm Higgs}\; p$ large enough to see?

\bu Will $pp$ elastic scattering at large $t$ be the same at LHC energies as
at ISR energies?

\bu {\bf Can we construct the theory of perturbative evolution at small $x$?}

\bu {\it We need to combine perturbative and nonperturbative concepts ---
evolution and analyticity.}

\bibliography{book}
\end{document}